# Comparative study of Deep Learning Models for Binary Classification on Combined Pulmonary Chest X-Ray dataset

line 1: 1st Shabbir Ahmed Shuvo
line 2:
line 3: *Offenburg University of Applied Sciences*
line 4: Offenburg, Germany
line 5: shuvo.shabbirahmed@gmail.com

line 1: Md Aminul Islam
line 2: *School of Computing and Technology*
line 3: *University of Gloucestershire*
line 4: Gloucester, UK
line 5: talukder.rana.13@gmail.com

line 1: 3rd Md. Mozammel Hoque
line 2:
line 3: *RNB Lab*
line 4: Dhaka, Bangladesh
line 5: mozammel2030@gmail.com

line 1: 4th Rejwan Bin Sulaiman
line 2:
line 3: *Northumbria University*
line 4: London, UK
line 5: rejwan.sulaiman@northumbria.ac.uk

***Abstract*—CNN-based deep learning models for disease detection have become popular recently. We compared the binary classification performance of eight prominent deep learning models: DenseNet 121, DenseNet 169, DenseNet 201, EffecientNet b0, EffecientNet lite4, GoogleNet, MobileNet and ResNet18 for their binary classification performance on combined Pulmonary Chest X-rays dataset.Despite the widespread application in different fields in medical images, there remains a knowledge gap in determining their relative performance when applied to the same dataset, a gap this study aimed to address.The dataset combined Shenzhen, China (CH) and Montgomery, USA (MC) data. We trained our model for binary classification, calculated different parameters of the mentioned models, and compared them. The models were trained to keep in mind all following the same training parameters to maintain a controlled comparison environment.End of the study, we found a distinct difference in performance among the other models when applied to the pulmonary chest X-ray image dataset, where DenseNet169 performed with 89.38% and MobileNet with 92.2% precision.**

**Keywords—*Pulmonary, Deep Learning, Tuberculosis, Disease detection, X-ray***

## I. Introduction

Tuberculosis (TB) continues to pose a significant threat to global public health, affecting a considerable number of individuals worldwide. According to data presented by the World Health Organisation (WHO), it is anticipated that the prevalence of tuberculosis would see a rise to around 10.6 million individuals in the year 2021, indicating an increase from the 10.1 million cases documented in the preceding year of 2020. Unfortunately, in the year 2021, there was a significant number of fatalities, totaling 1.6 million, that were due to tuberculosis. Among these cases, it was observed that 187,000 individuals were HIV-positive. There has been a notable rise compared to the preceding year, 2020, wherein there were 1.5 million fatalities documented, encompassing 214,000 individuals who tested positive for HIV [1]. Despite significant progress in efforts to reduce tuberculosis (TB), it continues to be the leading cause of mortality caused by a single infectious agent, surpassing the burden of HIV/AIDS [2]. This alarming toll highlights the urgent need for accurate and timely TB diagnosis to improve patient outcomes and reduce transmission rates. Pulmonary chest X-rays have long been a cornerstone in TB diagnosis, offering a non-invasive method to visualize lung abnormalities associated with the disease [3]. However, manual interpretation of these X-rays by radiologists can be time-consuming and subjective, leading to variations in diagnostic accuracy and potential delays in treatment initiation. Moreover, with the increasing number of TB cases worldwide, the burden on healthcare systems to handle this high volume of radiological data poses additional challenges in providing prompt and accurate diagnoses.

In recent years, researchers have turned to machine learning as a potential solution to enhance TB diagnosis from pulmonary chest X-rays. Machine learning algorithms can learn patterns from vast datasets, and when applied to medical imaging, they have the potential to aid radiologists in detecting subtle features indicative of TB with high accuracy[4-7]. Machine learning can expedite diagnosis, facilitate early detection, and improve overall patient outcomes by automating and optimizing the diagnostic process.

In this study, we conducted a comprehensive analysis by combining two diverse datasets, namely the Schenzen and Montgomery datasets, to create a larger and more robust dataset for tuberculosis (TB) diagnosis. Using this enriched dataset, we performed a comparative performance evaluation of eight deep transfer learning models: DenseNet121, DenseNet169, DenseNet201, EfficientNet B0, EfficientNet Lite4, GoogleNet, MobileNet, and ResNet18.

Our objective was to identify the most effective models for diagnosing TB accurately. We employed various performance metrics to assess the models' capabilities to achieve this. Through rigorous evaluation, we ranked the models based on their performance, presenting a clear hierarchy from the best-performing to the lowest. Furthermore, this study delves into the reasons behind the varying responses of these models. By analyzing the strengths and weaknesses of each model, we gained valuable insights into their suitability for TB diagnosis. The main contributions of our research are as follows:

- Fusion Dataset: By combining two datasets from two geographical regions, we created a larger and more representative dataset, improving our analysis's generalization and robustness.
- Comparative Performance Analysis: We rigorously compared eight deep transfer learning models,

providing a comprehensive understanding of their effectiveness in TB diagnosis.

- Performance Ranking: Our study presents a clear ranking of the models, helping clinicians and researchers identify the most promising models for practical implementation.
- Insightful Analysis: By examining the reasons behind the models' performance, we offer valuable guidance for model selection and potential areas of improvement.

Overall, our research contributes to advancing TB diagnosis using deep transfer learning techniques, ultimately supporting efforts to combat this global health challenge.

Section 2 of this article represents the literature review, and Section 4 depicts the materials and methodology of this study. The result analysis and discussion are explained in section 5. Section 4 points out the conclusion and future work of this study.

## II. LITERATURE REVIEW

Tuberculosis is a global threat. Multidrug-resistant bacteria and opportunistic infections in immunocompromised HIV/AIDS patients have made tuberculosis diagnosis harder. Untreated TB patients have high mortality rates. Diagnostics use century-old methods. Usually slow and unreliable. To reduce disease prevalence, the authors propose an automated method for detecting tuberculosis in traditional posteroanterior chest radiographs. Graph cut segmentation extracts the lung region first. They compute texture and form data for this lung region to binary classify X-rays as normal or aberrant. Their approach is tested using two datasets: one from the county's health department's tuberculosis control program in the US and one from Shenzhen Hospital in China. Field-ready tuberculosis screening computer-aided diagnostic system performs like human specialists. The first set has an area under the ROC curve (AUC) of 87% (78.3% accuracy) and the second set of 90% (84% accuracy). They initially compare their system to radiologists. Radiologists' false positive rate is half that of their technology, and their accuracy is 82% [8].

Deep learning algorithms have improved radiological picture anomaly identification, paving the path for its implementation in CAD systems. However, CAD systems for pulmonary tuberculosis (TB) diagnosis lack high-quality training data, adequate quantity and variety, and fine-region annotations [28].

The process of diagnosing digital chest X-rays (CXR) requires the detection of lung areas. This is achieved by a robust lung segmentation method that utilises nonrigid registration and image retrieval techniques. The method employs patient-specific adaptive lung models to accurately identify the boundaries of the lungs. The accuracy rates achieved on two chest X-ray (CXR) datasets [17], [18], and [19] from Montgomery County, Maryland, United States of America, and India, respectively, were 94.1% and 91.7%. The computation of lung area symmetry was performed by the authors through the use of multi-scale shape characteristics. This involved the incorporation of both local and global representations of the lung regions. Additionally, edge- and texture-based features were considered, taking into account the interior content of the regions. The researchers have provided evidence to support the suitability of their feature representation for the purpose of chest X-ray screening to detect pulmonary problems. The researchers achieved encouraging results on two benchmark datasets of chest X-ray (CXR) images from the National Institutes of Health (NIH) and India by employing a voting-based ensemble approach that combines three distinct classifiers: random forest (RF), multilayer perception (MLP) neural networks, and Bayesian network (BN). The greatest accuracy for detecting abnormalities (ACC) was found to be 91.0%, with an area under the receiver operating characteristic curve (AUC) of 0.96. This was achieved by the cross-population test, which yielded a maximum abnormality detection accuracy (ACC) of 89.0% and an AUC of 0.96. In this scholarly investigation, a modified version of the AlexNet architecture, referred to as MAN, is introduced for the purpose of detecting lung abnormalities in biomedical imaging. The proposed approach involves the comparison of lung CT scans with chest X-rays [21].

During the preliminary assessment, the medical artificial neural network (MAN) categorises the chest X-ray images into two classes: normal and pneumonia. The deep learning (DL) strategy recommended in this study demonstrates a higher level of accuracy (>96%) compared to other DL strategies that were evaluated. The classification of lung CT images into malignant or benign categories is determined by the analysis of MAN architecture, both with and without the inclusion of EFT.

The classification accuracy of the recommended Multiple Attention Network (MAN) with Support Vector Machine (SVM) classifier, which is 86.45%, is noticeably lower compared to the classification accuracy of a comparable Deep Learning (DL) framework with EfficientNet Transfer Learning (EFT), which exceeds 97%. This comparison highlights that the MAN framework exhibits satisfactory performance when applied to picture datasets. The data utilised in this study was obtained from the CXR arm of the PLCO dataset, which is a comprehensive lung cancer screening trial. The dataset consisted of a vast collection of 198,000,000 chest X-rays (CXRs) that were annotated with abnormal information based on the images. These CXRs were gathered from various clinical sites around the United States [23]. The terms Hierarchical Label Conditional Probability (HLCP) and Unconditional Probability generated from the Chain Rule are denoted as HLCP and HLUP, as mentioned in reference [23].

The HLUP (High-Level Understanding of Patterns) has been modified by incorporating the Cross-Entropy (CE) loss function and applying the chain rule. Table 1: Summary Table of Relevant Works for Pulmonary CXR Detection.

Table 1: Summary Table of Relevant Works for Pulmonary CXR Detection.

| Refs | Dataset | Models | Accuracy (%) | Note, Year |
|---|---|---|---|---|
| [8] | Montgomery County | SVM Classifier | 78.3 | AUC is 86.9% and sensitivity is 95%, 2014 |
| [8] | Shenzen Hospital | SVM (Linear)<br>SVM (PK)<br>SVM (RBF Karnels)<br>NN<br>ADT<br>LLR | 82.5<br>76.4<br>76.4<br>80.7<br>82.6<br>84.1 | AUC is approximately 88% for set A, and AUC is 88.5% for set B, 2014 |
| [10] | JSRT Database<br>Montgomery County<br>India | robust lung boundary detection method | 95.4<br>94.1<br>91.7 | X-rays of analog imaging system;<br>Montgomery and India CXR datasets acquired using digital scanner, 2014 |
| [15] | Chest X-ray 14 | Knowledge Distillation Deep Learning | 82.6(Average) | DenseNet-121 AUC80.97%, ResNet-152 (79.01%),<br>VGG-19 (76.17%), ResNet-50 (71.66%), MobileNet-v1(67.10%), and ResNet-32 (66.05%), 2020 |
| [16] | MC (USA), CH (China), IN(India) | BN<br>RF<br>MLP | 95 for RF, CH | All cases combination of 3 classifiers give highest accuracy, 2017 |
| [20] | Chest X-Ray<br>LIDC-IDRI database | Modified AlexNet (MAN)-SVM | 97.27 | Handicrafted and learned features give maximum accuracy over VGG 16, 19, AlexNet, RestNet50, MAN-SoftMax, MAN-KNN, MAN-RF,2020 |
| [22] | CXR arms PLCO dataset | HLUP-finetune | 88.7 | It performs best over BR-leaf, BR-all, HLUP, HLCP, 2020 |
| [24] | PLCO(NCI) +Chest X-ray(NIH) | Multi-task Learning | 88.3 | Lung and heart segmentations, spatial labels, adaptive normalization strategy, Re-labelling (94.5), 2019 |
| [25] | CXR, Municipal Hospital Czech<br>Republic | DLAD | 86.6 | Carebot AI CXR v1.22, 2023 |
| [26] | Six different dataset | CovTbPnNet | 99.76 (TB) | proposed model CovTbPnNet used for TB, Pneumonia, Covid 19, 2022 |
| [27] | bounding box dataset from CXR | RetinaNet | 77 | Precision 0.89, recall 0.57, specificity 0.94 , 2020 |
| [30] | 1007 posteroanterior chest radiographs | AlexNet, GoogLeNet | 99 | DCNN, 2017 |
| [32] | MC, Shenzhen, JSRT | TransUNet | 98.36 | Other used model: ARSeg, TransM, MedicalTransformer(MedT), TransUNet, UNeXt, 2023 |

## III. Materials and Methodology

### 3.1. Dataset Collection

Two publicly available chest X-ray datasets have been produced for the purpose of diagnosing tuberculosis. These datasets are known as the Shenzhen dataset [32] and the Montgomery country dataset [33]. Shenzhen serves as the primary digital picture database for tuberculosis. The collaborative effort between the National Library of Medicine in Maryland, United States, and the Shenzhen No. 3 People's Hospital, Guangdong Medical College in Shenzhen, China, results in the production of this resource. The dataset comprises 662 instances, with a class distribution consisting of 336 instances exhibiting symptoms of tuberculosis and 326 instances classified as normal. The Montgomery County X-ray database comprises X-ray pictures acquired from the tuberculosis control program administered by the Department of Health and Human Services in Montgomery County, Maryland, United States of America. The dataset has a total of 138 thoracic X-rays, with a class distribution of 80 X-rays classified as normal and 58 X-rays exhibiting symptoms indicative of tuberculosis. It is imperative to acknowledge that both datasets are focused on the identification of tuberculosis in chest X-ray pictures. The "China Set - The Shenzhen set" comprises a greater quantity of X-ray images in comparison to the "Montgomery County X-ray Set," which, although fewer in number, still has a significant number of photos. Each data set has both instances of normal cases and instances of tuberculosis-affected patients, rendering them suitable for the purpose of training and assessing machine-learning models designed for tuberculosis detection.

We have merged the two datasets with some preprocessing. This has ensured that the final dataset is consistent, balanced, and ready for use in training a tuberculosis detection model. Proper data preparation and management are crucial to achieving accurate and reliable results when applying machine-learning techniques to medical image analysis tasks [34].

### 3.2. Dataset pre-processing

In training our deep learning models, we utilized a series of systematic transformations, carefully aligned to fit our specific needs, given the dataset's characteristics.

The first important step for our dataset preprocessing was to prepare the labels for our dataset. Our dataset comprises X-ray images in a folder and labels in another folder in text format. The labels were text files that contained the diagnosis in string. But the string contained in the labels varied from sample to sample. As we used binary classification for this work, we cleaned the labels first from unnecessary information for this study, and we labeled normal diagnosis as normal and diseased samples as TB (Tuberculosis). Then only we proceeded with the next steps in our data pre-processing.

As our dataset, composed of X-ray images, was already preprocessed, and converted into a universally recognized PNG image format, we could bypass extensive preprocessing steps. Nevertheless, we faced the challenge of a relatively small sample size for the training of deep learning models, a common concern in many machine learning applications.

To mitigate this challenge, we engaged the practice of Data Augmentation, specifically by employing 224x224 random resized crops of the X-ray images. This method effectively expanded our sample size and introduced a level of variability that would enable our model to generalize better to unseen data.

Alongside implementing random resized crops, we also applied a Random Horizontal Flip. This is a common practice for increasing the variability of the data further, enhancing the robustness of the model against different orientations of the same image.

Before the transformed images were inputted into our deep learning models, we converted them into tensors. This standard step transforms the image data into a numerical format that the deep learning model can efficiently process efficiently.

Finally, it's important to note that we used models pre-trained on ImageNet. To ensure consistency and transferability of learned features from the pre-trained models, we normalized our images using the specific ImageNet normalization parameters. These parameters are the mean value [0.485, 0.456, 0.406], and the standard deviation value [0.229, 0.224, 0.225].

Ultimately, we split our dataset into training, validation, and test set. After splitting, the training set contained 60% of the total data samples. The training and validation both contained 20% of the total data samples.

This comprehensive preprocessing and transformation approach allowed us to optimize the limited resources of our dataset while preparing it for an effective deep-learning training process.

3.3 Workflow and model preparation

In our work, we used a collection of strong, pre-built deep learning models that were chosen for their reliable feature extraction abilities developed on large image datasets. Due to their consistently higher performance, our models are often used in classification tasks. Our selected models are: DenseNet121, DenseNet169, DenseNet201, EfficientNet-b0, EfficientNet-Lite4, GoogLeNet and ResNet18.

In this study, we worked with binary classification, and we followed the workflow as shown in in Fig 1. As discussed before, we first selected eight pre-trained models trained on the imagenet dataset. As these models learned the general characteristics of standard images, so their model weight is a good starting point to start our model training. But as we are working with X-ray medical images with very different characteristics than standard images, we decided to train the full models without freezing any layers. As mentioned in the previous dataset preparation section, we split the total dataset on training, validation, and test set. Then we trained our model on the training set and evaluated on the validation set. At the end when the training was complete, we tested our model on the prepared test set. To implement this workflow we used a Python-based Pytorch deep learning library along with standard machine and deep learning tools such as sklearn, PIL (Python Imaging Library), matplotlib, seaborn, pandas, numpyb etc.

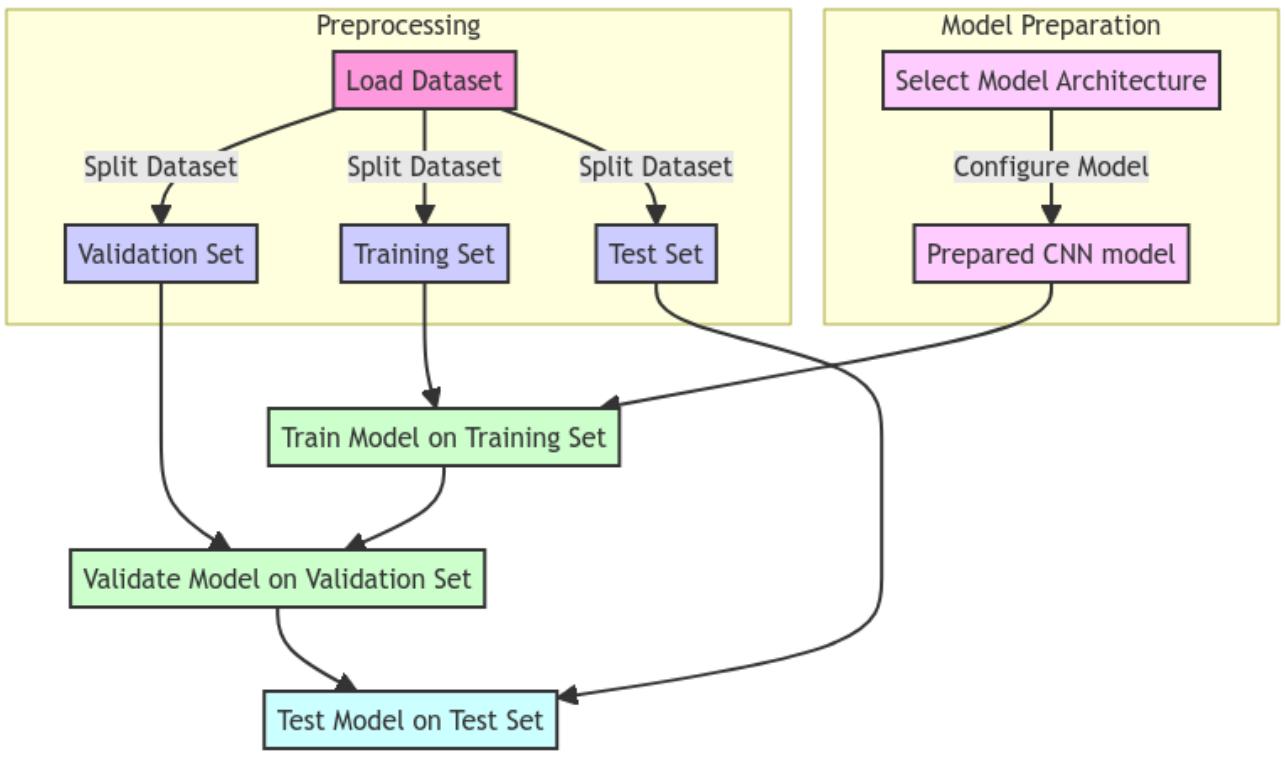

**Fig 1: Proposed workflow diagram**

Our work can be divided into four main stages namely:

1. Data preprocessing
2. Model preparation
3. Model training
4. Model evaluation/testing

In figure 2, we can see the the steps involved data preprocessing, model preparation, training and evaluation processes. The data preprocessing stage involves data loading, data samples resizing, data augmentation and data normalization of the data samples in the dataset. In model preparation we loaded the pre trained models then modified the fully connected classification layer for binary classification. Then we compiled the model for the training. The training process contains forward pass, backward pass, updating weights and validation. The forward pass process contains of forward propagation and loss computation. We used crossentropy loss function for loss calculation. In the backward pass the is performed to calculate the gradients of the loss with respect to the model's parameters. In update weights step the optimizer uses these gradients to update the model's parameters. We used Stochastic Gradient Descent (SGD) optimizer with initial learning rate 0.001 and momentum 0.9. For optimal model training we additionally used learning rate scheduler(StepLR) in the training process. For learning rate scheduler settings we used step_size=7, gamma=0.1 where step_size represents the number of epochs after which the learning rate is multiplied by multiplication factor gamma. We trained each model for 100 epochs. During training we saved the best model weights for evaluation in the next step. During evaluation we loaded the best model weight saved during training. Then we tested the model with the test dataset and calculated different relevant metrics and plotted relevant diagrams which will be discussed in the results section.

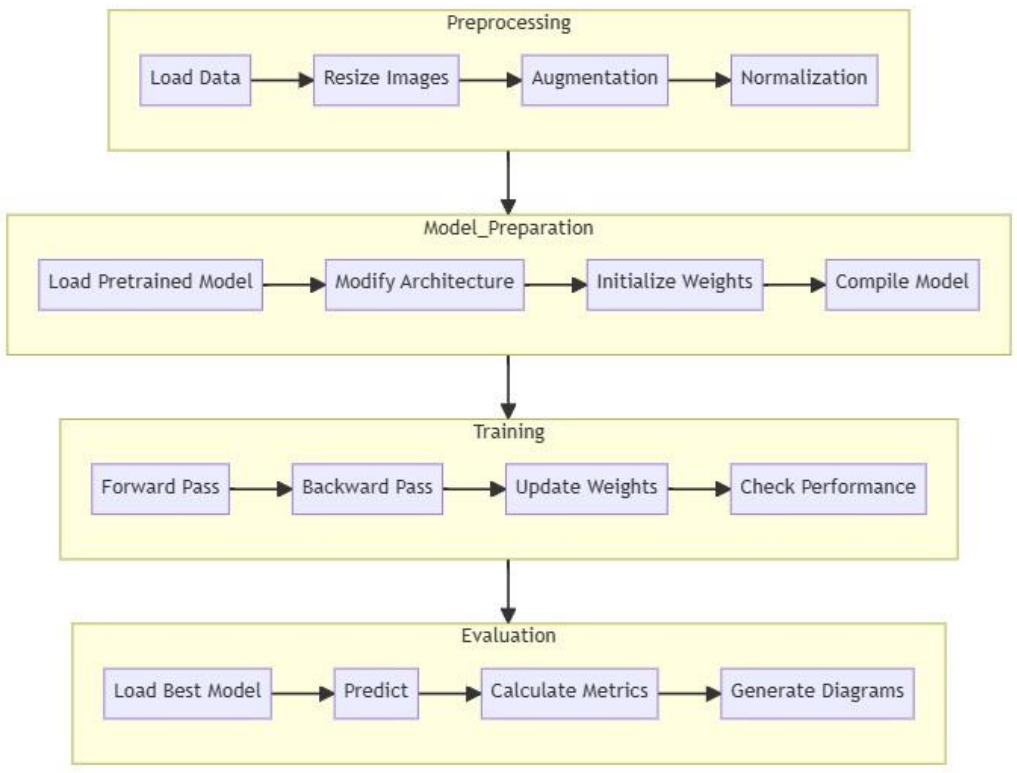

**Fig 2: Model training and evaluation block diagram**

3.4 Model Evaluation

We have chosen appropriate evaluation metrics to assess the model's performance. The used metrics for binary classification tasks like tuberculosis detection are:

- **Accuracy:** The proportion of correctly classified samples.

$$Accuracy = \frac{True\ positive + True\ Negative}{Total\ number\ of\ cases}$$

- **Precision:** The ability of the model to correctly identify true positive cases out of all predicted positive cases.

$$Precision = \frac{True\ positives}{True\ positives + False\ positives}$$

- **Recall (Sensitivity or True Positive Rate)**: The ability of the model to correctly identify true positive cases out of all actual positive cases.

$$Recall = \frac{True\ positives}{True\ positives + False\ Negatives}$$

- **F1 Score**: The harmonic mean of precision and recall, providing a balanced measure between the two.

$$F1\ score = \frac{2 * Precision * Recall}{Precision + Recall}$$

- **Confusion Matrix**: Construct a confusion matrix to visualize the model's performance and gain insights into true positives, false positives, true negatives, and false negatives.
- **Receiver Operating Characteristic (ROC) Curve**: Plot the ROC curve and calculate the Area Under the Curve (AUC) to evaluate the model's performance across different probability thresholds.
- **Precision-Recall Curve**: Plot the precision-recall curve to examine the trade-off between precision and recall at various probability thresholds.

It's crucial to emphasize that medical applications require thorough validation and consultation with domain experts to ensure the model's reliability and safety before deploying it in real-world settings. The chosen evaluation metrics and procedures have aligned with our objectives and requirements of the tuberculosis detection task using chest X-ray images.

## IV. RESULT ANALYSIS AND DISCUSSION

The following table highlights the performance of different convolutional neural network (CNN) architectures evaluated on our binary classification task. Performance metrics such as Accuracy, Precision, Recall, F1 Score, and ROC AUC are used to understand the strengths and weaknesses of each model.

Table 2: Summary of performance metrics of CNN models

| Model Name | Accuracy (%) | Precision (%) | Recall (%) | F1 score (%) | ROC AUC (%) |
|---|---|---|---|---|---|
| DenseNet121 | 84.375 | 89.706 | 77.215 | 82.992 | 89.717 |
| DenseNet169 | 89.375 | 89.744 | 88.608 | 89.172 | 92.202 |
| DenseNet201 | 84.375 | 90.909 | 75.949 | 82.759 | 90.045 |
| EffecientNet b0 | 79.375 | 84.848 | 70.886 | 77.241 | 87.436 |
| EffecientNet lite4 | 78.125 | 80.556 | 73.418 | 76.821 | 85.045 |
| GoogleNet | 80.625 | 88.710 | 69.620 | 78.014 | 87.654 |
| MobileNet | 85.625 | 92.424 | 77.215 | 84.138 | 91.858 |
| ResNet18 | 85.000 | 87.671 | 81.013 | 84.211 | 89.905 |

Analyzing the given results, the following observations can be made:

DenseNet169 performed the best in terms of accuracy (89.375%) and ROC AUC (92.202%), suggesting that it was the most effective model at classifying the given data correctly, as well as distinguishing between classes.

DenseNet201 and MobileNet showed high precision (90.909% and 92.424% respectively), meaning that they were quite adept at correctly predicting the positive class and minimizing false positives.

ResNet18 had the highest recall (81.013%) of all models, indicating that it was the best at identifying all true positives in the data.

DenseNet169 also had the highest F1 score (89.172%), which suggests it had a good balance between precision and recall, and could maintain robust performance in different conditions.

It is worth noting that even though EfficientNet b0 and EfficientNet lite4 had the lowest accuracy scores, they still maintained respectable ROC AUC scores (87.436% and 85.045% respectively). This might indicate their relatively better performance in balancing true positive and false positive rates across different thresholds.

Finally considering overall performance on our specific dataset, and the importance of precision and recall in medical diagnostics it can be said that, DenseNet169 is an effective model for TB detection for our dataset with X-ray images. In the figure below we can see the training - validation loss plot along with confusion metrics and ROC curve for DenseNet169.

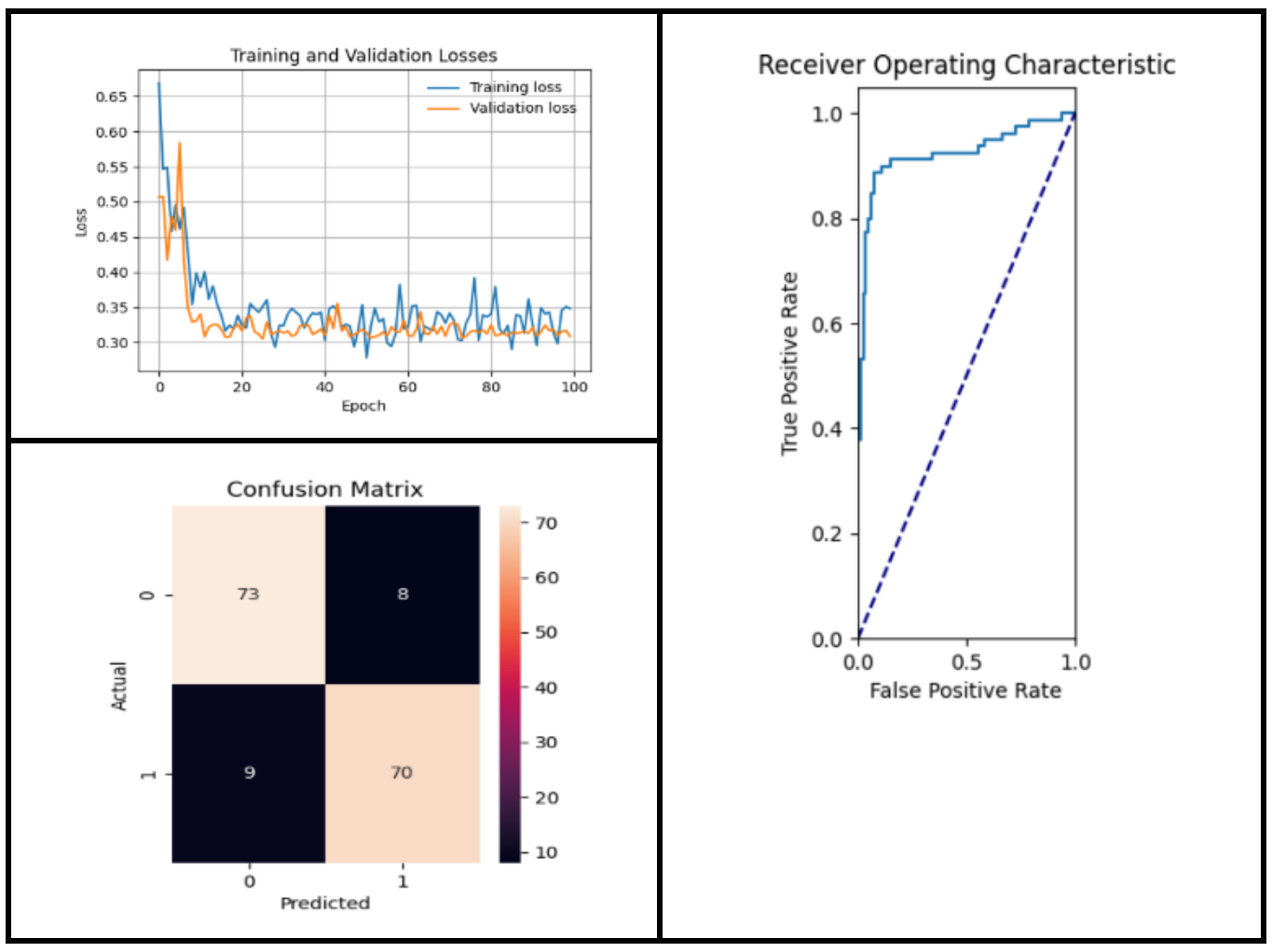

**Fig 3: Training-validation loss plot, confusion metrics and ROC curve for DenseNet169**

## V. CONCLUSION AND FUTURE WORK

The performance differences among the models may be due to factors like architecture depth, number of parameters, receptive field size, and the type and arrangement of layers. Each architecture has its own strengths and weaknesses, making it more or less suitable for different tasks or data distributions.

This research has achieved our target of achieving higher accuracy using less complex pre-processing techniques and less computational power. DenseNet169 has demonstrated the highest values in four out of five performance matrices: accuracy 89.375%, Precision 89.744%, F1 score 89.172%, and ROC area under the curve 92.202%. Regarding Recall, we got the highest value from MobileNet is 92.424% (>88.608% of DenseNet169 and >90.909% of DenseNet201). We suggest using this proposed model along with manual crosschecks by the clinicians before the final decision of surgery or serious medical operation. Our research is a clear indication of readily available image dataset (MC and CH) usage for TB detection, which can be turned into a device for developing and under-developed countries to automatize the health sector. There is a scope for future researchers to improve in gaining 100% accuracy.